\documentclass{article}           %% ceci est un commentaire (apres le caractere %)
\usepackage[T1]{fontenc}          %% permet d'utiliser les caractères accentués
\usepackage{graphicx}      %% permet d'importer des graphiques au format .EPS (postscript)
\usepackage{fancybox}		   %% package utiliser pour avoir un encadré 3D des images
\usepackage{makeidx}              %% permet de générer un index automatiquement
\usepackage{graphicx}
\usepackage{authblk}

\usepackage[T1]{fontenc}
\usepackage[utf8]{inputenc}
\usepackage{pgf}
\usepackage{tikz}

\begin{document}
\title{High-harmonics of harmonics of a fiber laser: a milliwatt XUV ultrashort source}

\author[1,*]{ A. Comby}
\author[1]{D. Descamps.}
\author[1]{S. Beauvarlet.}
\author[2]{A. Gonzalez}
\author[2]{F. Guichard}
\author[1]{S. Petit}
\author[2]{Y. Zaouter}
\author[1]{Y. Mairesse}
\affil[1]{\textit{Universit\'e de Bordeaux - CNRS - CEA, CELIA, UMR5107, F33405 Talence, France}}
\affil[2]{\textit{Amplitude Laser Group, 33600 Pessac, France}}
\affil[*]{\textit{Corresponding author: antoine.comby@u-bordeaux.fr}}
\date{} 

\maketitle
    
\subsection*{Abstract}
Recent progresses in femtosecond ytterbium-doped fiber laser technology are opening new perspectives in strong field physics and attosecond science. High-order harmonic generation from these systems is particularly interesting because it provides high flux beams of ultrashort extreme ultraviolet radiation. A lot of efforts have been devoted to optimize the macroscopic generation parameters. Here we investigate the possibility of enhancing the single-atom response by producing high-order harmonics from the second, third and fourth harmonics of a turnkey femtosecond Yb-fiber laser providing at 166 kHz pulses of 50 W - 130 fs at 1030 nm. We show that the harmonic efficiency is optimal when the process is driven by the third harmonic, producing $5\pm1 \times 10^{14}$  photon/s at 18 eV, which corresponds to 1.5$\pm$0.3 mW average power.

\section*{Introduction}

High-order Harmonic Generation (HHG) is an unique tool to produce ultrashort pulses of spatially coherent extreme ultraviolet (XUV) radiation with table top systems \cite{calegari_advances_2016}. HHG occurs when atoms interact with an intense laser pulse, in the few $10^{14}$ W/cm$^2$ intensity range. Within one optical cycle, the strong laser field ionizes the atom, accelerates the freed electron wavepacket and drives it back to recombine with the parent ion, emitting an attosecond burst of XUV radiation \cite{corkum_plasma_1993,paul_observation_2001}. This process repeats periodically every half laser cycle, resulting in a broad spectrum consisting of odd harmonics of the driving laser frequency, extending until a cutoff frequency determined by the laser intensity and its wavelength \cite{lewenstein_theory_1994}.  

In the past 10 years, the emergence of compact, high power, high repetition rate femtosecond lasers based on ytterbium-doped fiber amplifiers (YDFA) has opened new perspective of applications for HHG. In 2009, pioneering experiments demonstrated the generation of high-order harmonics with repetition rates up to the MHz range \cite{boullet_high-order_2009}. Since then, a lot of efforts have been dedicated to optimize the XUV photon flux of these sources \cite{cabasse_optimization_2012,hadrich_single-pass_2016,heyl_introduction_2017}. Indeed, producing a high average XUV photon flux with a high repetition rate is optimal for experiments based on coincidence detection (e.g. COLTRIMS \cite{dorner_cold_2000,gagnon_time-resolved_2008,rothhardt_high-repetition-rate_2016}) or surface photoemission measurements \cite{puppin_time-_2019}. 

The low energy per pulse of YDFA lasers, compared to Ti:Sa lasers standardly used in HHG experiments, imposes tighter focusing conditions. This was initially thought to be detrimental for HHG. However important works recenlty demonstrated that the experimental parameters (pressure, medium length, focal length) could be scaled to keep good phase matching conditions and optimize the HHG efficiency \cite{heyl_high-order_2012,rothhardt_absorption-limited_2014,heyl_scale-invariant_2016}. %We recently implemented a new technique to measure the full density profile of the gas jet \cite{comby_absolute_2018,horke_characterizing_2017}, confirming the possibility to achieve good phase-matching in the generation process. 

The second important limitation of YDFA lasers is their pulse duration, which is typically one order of magnitude longer than Ti:Sa lasers. This restricts the effective intensity that can be used for HHG before the generating medium is completely ionized, and thus the extension of the high-harmonic cutoff. Furthermore, ionization induces dispersion of the fundamental and XUV beams, which can lead to situations where the phase matching conditions are fulfilled only during a short part of the laser pulse duration (transient phase matching \cite{kazamias_pressure-induced_2011,heyl_introduction_2017}), limiting the conversion efficiency. Reducing the pulse duration of YDFA lasers is thus the subject of many works, by using single or double postcompression stages in hollow core fibers \cite{rothhardt_53w_2014,hadrich_energetic_2016,lavenu_high-energy_2017} and/or in cavities  \cite{lavenu_nonlinear_2018,ueffing_nonlinear_2018,kaumanns_multipass_2018}. The highest reported XUV flux from YDFA lasers indeed used such schemes, reaching 832 $\pm$ 204 $\mu$W with postcompression \cite{klas_table-top_2016} and $\sim$ 2 mW from cavity HHG \cite{porat_phase-matched_2018}. 

Here we show that postcompression or cavity enhancement are not necessary to produce mW-class XUV sources from YDFA lasers, leading to a significant reduction of experimental complexity. The HHG conversion efficiency is known to depend dramatically on the generating laser frequency $\omega_L$ \cite{balcou_optimizing_1992,shiner_wavelength_2009,lai_wavelength_2013,marceau_wavelength_2017,wang_bright_2015,popmintchev_ultraviolet_2015}, 
the process being more efficient when driven by low frequency lasers. This enhancement is mostly caused at the single-atom level by a lower spreading of the electronic wavepacket in the continuum, which leads to a higher recombining current with the parent ion. While all investigations conducted so far confirmed a general trend of XUV flux increase with increasing generating frequency $\omega_L$, the scaling laws measured in different experimental conditions were very different: $\omega_L^{6-7}$ \cite{shiner_wavelength_2009} in the mid-IR range, and $\omega_L^{4-5}$ \cite{lai_wavelength_2013,marceau_wavelength_2017}, $\omega_L^{6} $ \cite{wang_bright_2015} or $\omega_L^{8}$ \cite{popmintchev_ultraviolet_2015} in the visible-UV range.

In this article we investigate the generation of high-order harmonics from the second, third and fourth harmonics of a femtosecond YDFA laser system. The resulting source is extremely simple and versatile, offering the possibility to generate spectra with different energy spacings of the XUV comb, and in different spectral ranges. We observe a strong enhancement of the harmonic conversion efficiency in the low energy region of the spectrum when HHG is driven by the third harmonic, enabling the production of $4.4 \times 10^ {14}$ photons/s at 18 eV,  an average power of 1.5 mW and a conversion efficiency of $1.6 \times10^{-4}$.

\section{Generation of the 515 nmn, 343nm and 257 nm beams.}

\begin{figure}[htbp]
\centering
\fbox{\includegraphics[width=\linewidth]{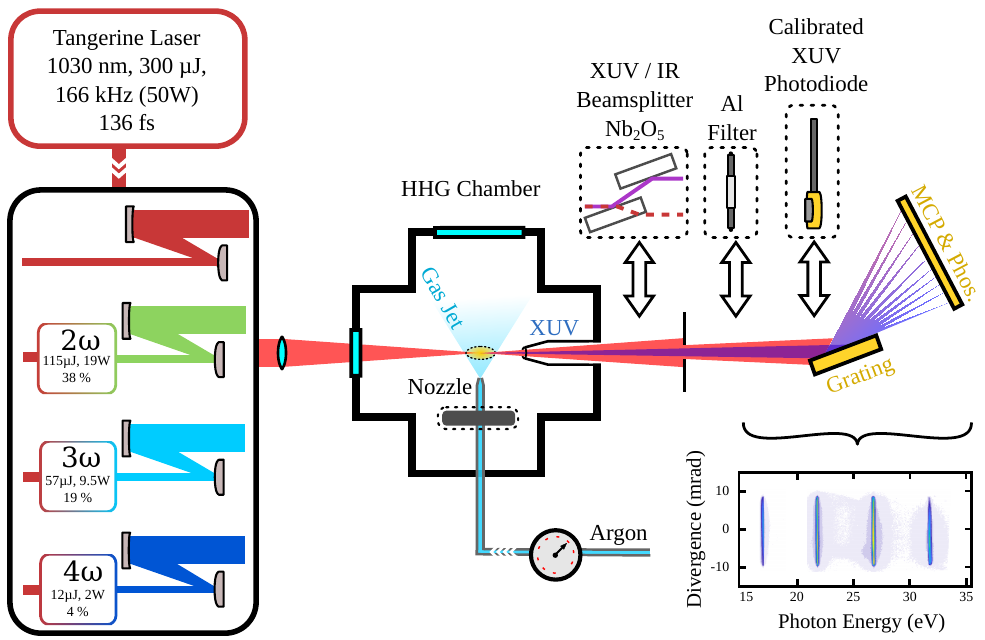}}
\caption{Scheme of the experimental setup. We generate the harmonics (2nd, 3rd and 4th) of the YDFA laser frequency in BBO crystals and after dichroic mirrors and telescopes, the converted frequency laser beam is focused in argon gas jet. The flux of the XUV spectrum is measured thanks to a spatially resolved XUV spectrometer and a calibrated photodiode.
}
\label{fig1}
\end{figure}

The experimental setup is described in Fig. \ref{fig1}. We used the BlastBeat femtosecond laser system at CELIA, which consists of two synchronized 50 W Yb-doped fiber lasers (Tangerine, Amplitude Systemes \cite{lavenu_high-energy_2017}) delivering  130 fs pulses centered at 1030 nm (FWHM = 18.5 nm), at a repetition rate which can be continuously tuned between 166 kHz and 2 MHz.  For this  study, we used a single arm at 166 kHz and we converted the $\omega_L$ fundamental frequency to 2, 3 and 4$\omega_L$ (1030 to 515, 343, 257 nm) by using beta barium borate (BBO) crystals.

To produce the second harmonic  (2$\omega_L$), we directly sent the 1030 nm beam in a 1 mm thick type-I BBO crystal and used two dichroic mirrors to select the up-converted frequency. We reached a conversion efficiency of 38 $\%$, leading to an average power of 19 W and a pulse energy of 130 $\mu$J. The pulse duration of the 2$\omega_L$ beam, characterized using an home-made SHG-FROG, was 130 fs FWHM. 

For the third harmonic (3$\omega_L$), we used an in-line frequency conversion setup. 
A first type-I BBO crystal ($\theta$ = 23 $^{\circ}$) was used to frequency double a fraction of the fundamental $\omega_L$ beam. A type-I 0.75mm thick BBO crystal performed the sum-frequency mixing at 3$\omega_L$. After 3 dichroic mirrors (HT$\omega_L$-2$\omega_L$ /HR 3$\omega_L$), we obtained a maximum average power of 12 W at 3$\omega_L$ (72 $\mu$J, 24 $\%$ conversion efficiency). However, to avoid detrimental thermal effect in the second BBO crystal, we routinely limited the 3$\omega_L$ average power to 9.5 W (19 $\%$ efficiency). Non-linear optics calculations with SNLO estimate the pulse duration of the 3$\omega_L$ beam to be around 140 fs.

For the fourth harmonic (4$\omega_L$) generation, we used the 2$\omega_L$ generation setup describe above, and sent the 2$\omega_L$ beam in a 200 $\mu$m type-I BBO crystal. The thickness of the BBO crystal is here limited by the large difference of group velocities of the 2$\omega_L$ (515 nm) and 4$\omega_L$ (257.5 nm) in BBO, which prevents temporal overlap over long propagation lengths. It is also necessary to avoid heating of the BBO crystal due to two-photon absorption in the UV \cite{dubietis_two-photon_2000}. This was achieved by setting a -150/250 telescope on the 2$\omega_L$ before doubling it,  without any loss in the efficiency. We reached a 4 $\%$ conversion efficiency from the initial laser power, i.e. 12 $\mu$J per pulse at 166 kHz (2 W) with an expected pulse duration about 130 fs. Note that to avoid thermally-induced wavefront distortions, only UV-grade fused-silica or high quality CaF$_2$ optical components were used with the 4$\omega_L$ beam. 

\section{Spatial profiles of the driving laser beams}

The spatial quality of the laser beams is a crucial parameter in high-order harmonic generation. In order to investigate the possible wavefront distortions of the beams due to thermal effects induced by the high power laser source, we performed measurements of the spatial profiles of the fundamental and harmonic beams under conditions close to the optimal ones for HHG. 

\begin{figure}[htbp]
\centering
\fbox{\includegraphics[width=\linewidth]{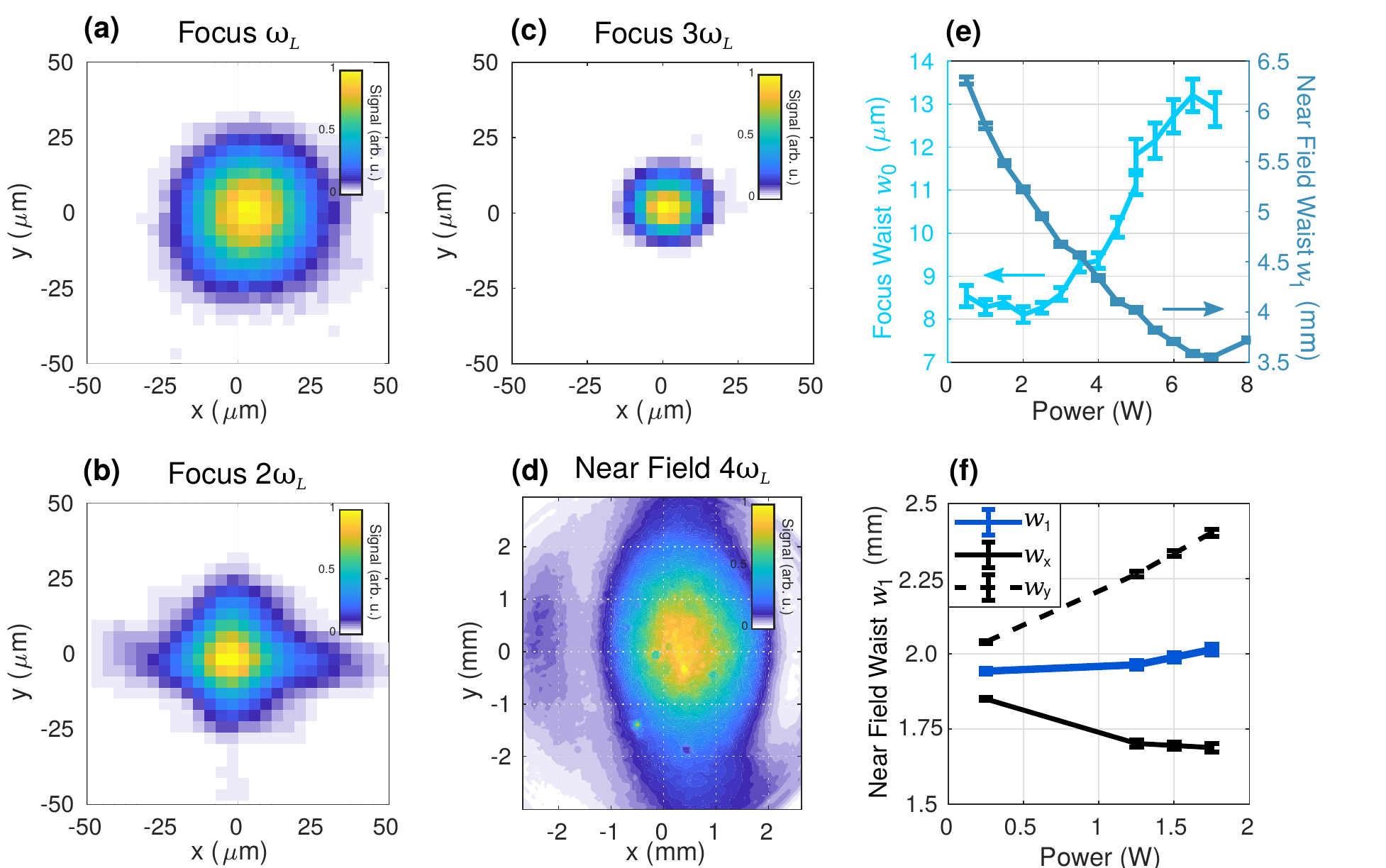}}
\caption{Measurement of the spatial profiles. (a) Focus of $\omega_L$ after -100 mm/300 mm telescope and f=300 mm at 50 W. 
(b) Focus of 2$\omega_L$ after -100 mm/300 telescope and f=300 mm at 19 W. 
(c) Focus of 3$\omega_L$ after -100 mm/500 mm telescope and f=300 mm at 7 W. 
(d) Near-Field of 4$\omega_L$ after -100/500 telescope at 1.75 W. 
(e) Beam waist measurement at 3$\omega_L$ after a -100 mm/500 mm telescope ($w_1$) and at focus ($w_0$) with f=300 mm. 
(f) Beam waist measurement at 4$\omega_L$ after a -100 mm/500 mm telescope ($w_1$) and its projection on the x ($w_x$) and y axis ($w_y$). 
}

\label{profils_2}
\end{figure}

The initial waist of the unfocused laser beam is about 1.3 mm (radius at $I_{max}$/e$^2$). In order to reach high intensities for HHG while keeping all the focusing optics outside the vacuum chamber, we used telescopes to increase the beam size before focusing it. We found out that whatever the wavelength used from 1030 nm to 257 nm, UV fused silica material introduces wavefront distorsion at high power of femtosecond pulses when beam size is close or below 1 millimeter. Therefore all-reflective telescopes based on spherical dielectric mirrors were implemented to increase beams size. For the fundamental and second harmonic beams, we used a combination of -100 mm/300 mm focal mirrors, followed by a f = 300 mm lens. The resulting focal spots, imaged by a CCD camera with a 3.75 $\mu$m resolution, are shown in Fig. \ref{profils_2} (a-b). The fundamental beam shows excellent focus quality, with a beam waist $w_0$ of 26 $\mu$m. The second harmonic has a waist of  20 $\mu$m and shows a slightly crossed shape, indicating some astigmatism because of the non-zero incidence angle of the mirrors in the telescope. We monitored the evolution of the fundamental and second harmonic focus as a function of the laser power, and found that they were independent of the power. 

We also measured the initial waist of the laser before focusing $w_1$ by sending directly the beam on the CCD. We then estimate the  deviation from the diffraction limit $DDL$ by applying the formula $DDL = \pi w_0 w_1 / (\lambda f)$, with $f$ the focal length, $\lambda$ the driving wavelength. $DDL$ should be equal to 1 for a perfect gaussian beam and increases if the wavefront is distorted. We obtain $DDL$ of $\sim 1/1.2$ in the horizontal/vertical direction at $\omega_L$ and $\sim 1.2$ at 2$\omega_L$ in both directions. Both measurements indicate good spatial beam quality.

We performed similar measurements for the 3$\omega_L$ beam, after a  -100 mm/500 mm dielectric mirror telescope and a focal lens of 300 mm. The spatial profile of the resulting focus at 8 W average power is shown in Fig. \ref{profils_2} (c). Even if the spatial beam profile seems nice, we found out that its size was changing with average power. We thus characterized the beam waist at focus $w_0$ and the beam waist before the focusing lens $w_1$ as a function of the average power of the 3$\omega_L$ beam (Fig. \ref{profils_2} (e)). The near field waist $w_1$ diminishes from to 6.5 mm to 3.5 mm when the 3$\omega_L$ power increases from 0.5 to 8 W. This is accompanied by an increase of the spot size at the focus of the lens, from 8.5 $\mu$m to 13 $\mu$m. This is probably due to thermal effects in the crystal. These measurements lead to a $DDL$ of 1.4 at high power.

Last, we measured the near-field profile of the the 4$\omega_L$ after a -100 mm /500 mm dielectric  mirror telescope (Fig. \ref{profils_2} (d)). The resolution of our UV camera was insufficient to measure the focus.  The beam becomes 20$\%$ elliptical at high power, but the geometric mean waist stays around 2.0 mm. The appearence of the elliptical shape is due to the temperature increase in the crystals due to a non-linear absorption of the 257 nm, which changes the angular acceptance of the phase-matching process.

\section{Spatially Resolved HHG Profile}

The experimental setup to produce the XUV light through HHG is described in Fig. \ref{fig1}. After dichroic mirrors and a telescope, we focused each laser harmonics on a characterized 250 $\mu$m thick argon gas jet \cite{comby_absolute_2018} and produced HHG.
 The XUV was characterized by a flat field XUV spectrometer, consisting of a 1200 grooves/mm holographic cylindrical grating with variable groove spacing (Shimadzu) and a set of dual micro-channel plates coupled to a P46 (fast) phosphor screen (Hamamatsu). A 12-bit cooled CCD camera (PCO) recorded the spatially-resolved harmonic spectrum.

\begin{figure}[htbp]
\centering
\fbox{\includegraphics[width=\linewidth]{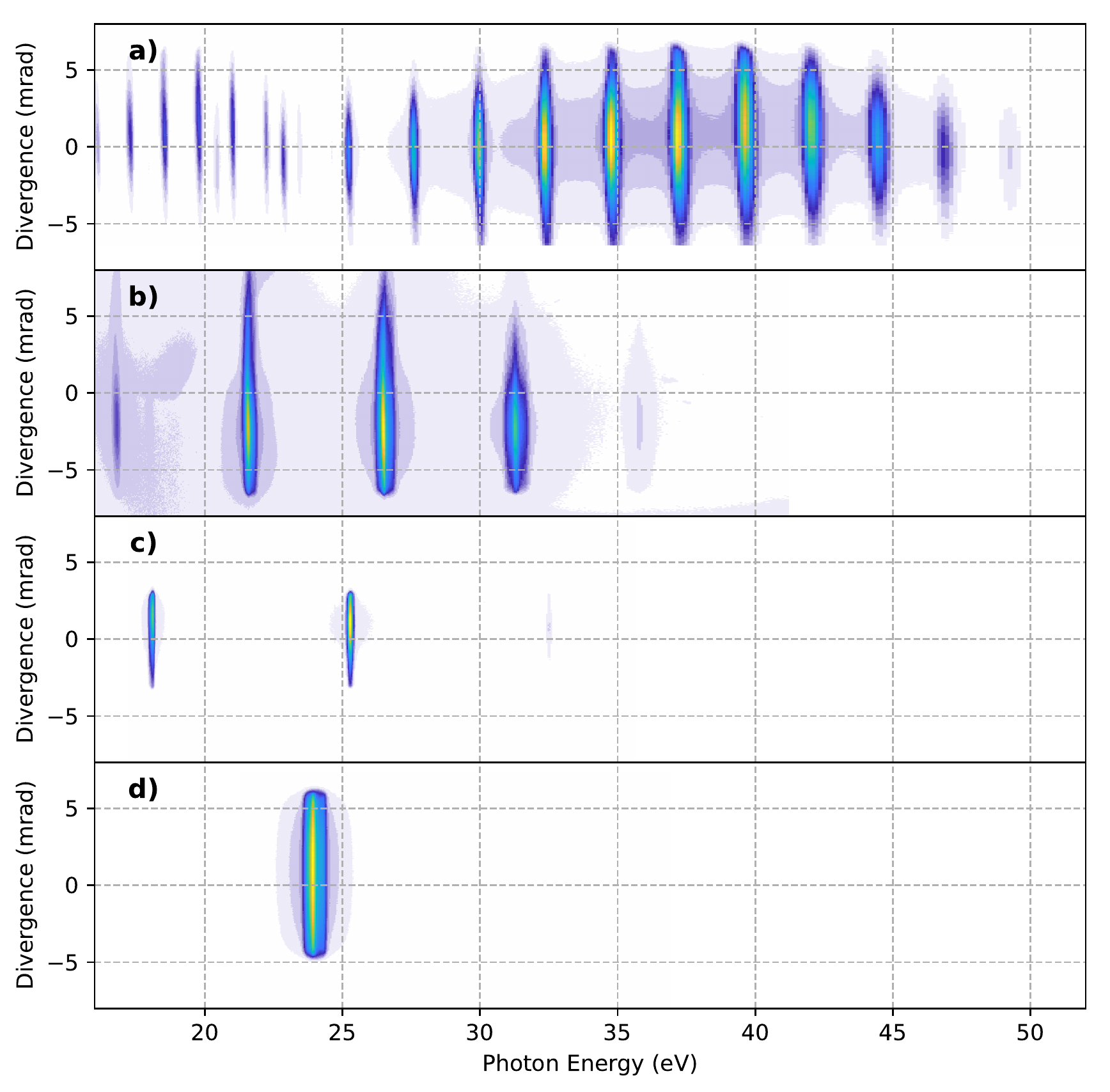}}
\caption{(a)-(d) Spatially resolved HHG spectrum generated in argon with 1,2,3 and 4$\omega_L$ driving laser frequency, using intensities of 1.5, 2.8, 3.4 and $1.6 \times 10^{14}$ W/cm$^2$. The peak density in the generating medium is $4.5\times10^{18}$ at./cm$^3$ at $\omega_L$ and $3.5 \times10^{18}$ at./cm$^3$ at 2,3 and 4 $\omega_L$. The spectra at 3 and 4 $\omega_L$ are recorded with an Al filter, otherwise the phosphor screen is excited from the UV scattering light in the spectrometer chamber.
}
\label{fig2}
\end{figure}

The spatially resolved high-order harmonics spectrum produced by the 1-4$\omega_L$  frequencies in optimized conditions are displayed in Fig. \ref{fig2}. Each spectrum is made of the odd harmonic of the laser frequency. The cutoff is, as expected, much higher at $\omega_L$ at 49.2 eV (H41) -- close to the Cooper minimum of argon \cite{cooper_photoionization_1962,higuet_high-order_2011} around 53 eV -- and reaches 36 eV for 2$\omega_L$ (H13), 25.2 eV at 3$\omega_L$ (H7) and 24 eV at 4$\omega_L$ (H5). The spectral width of the harmonics in the plateau region is too narrow to be resolved by our spectrometer, and is expected to be around 30 meV. Spatially, we observe a up/down clipping due to a slit ensuring differential pumping outside of the HHG chamber and occluding a large part of the laser power but we do not observe any other strong features. The excellent spatial beam quality of the high-order harmonics produced from YDFA laser demonstrates the potential of the source for imaging purposes.

\section{HHG Flux Measurements}

\begin{figure}[htbp]
\centering
\fbox{\includegraphics[width=\linewidth]{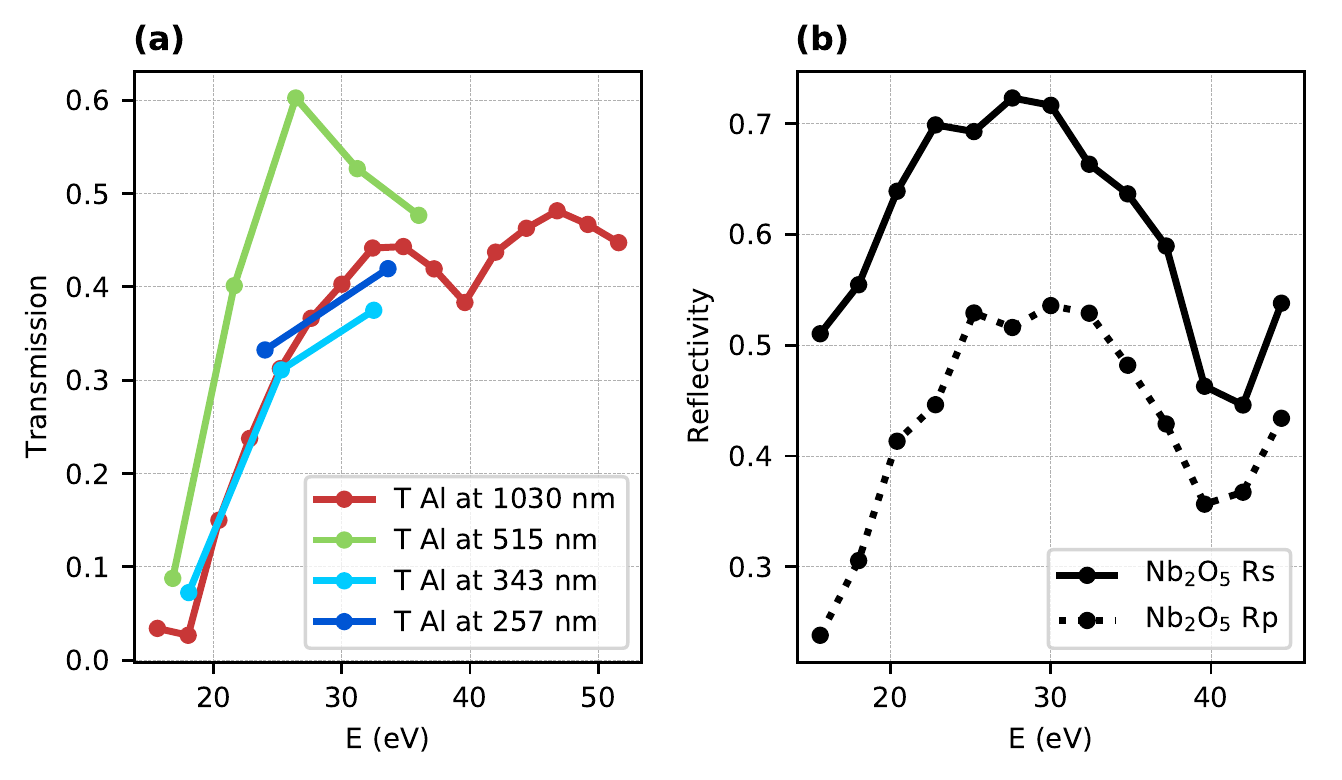}}
\caption{(a) Transmission of the 157 nm Al filters measured using different driving wavelengths. The filters used in each configuration are different. 
(b) Reflection rate of one Nb$_2$O$_5$ mirror in s- and p- polarization at $\sim 70^{\circ}$ incidence angle.
}
\label{fig5}
\end{figure}

To measure the HHG photon flux, we used an XUV photodiode calibrated by the Physikalisch-Technische Bundesanstalt Berlin, in front of a 157 nm thick Al filter whose transmission was measured directly with our XUV spectrometer. The measured filter transmission for each harmonic is plotted in Fig. \ref{fig5} (a). Each set of data corresponds to a different Al filter. The transmission maximizes around 45$\%$ in the 32-50 eV range, significanly below the expected 70$\%$ from CXRO \cite{henke_x-ray_1993}. This is the signature of oxydation of the Al filter where Al$_2$O$_3$ significant decreases the overall filter transmission. Different levels of transmissions are measured around 33 eV, illustrating the different level of oxydation of the different filters. The low energy decay of the transmission occurs above the expected 20 eV cutoff. This was observed in previous works, and interpreted as the result of multiple reflections of the XUV light in the filter (see Supplementary Information in \cite{wang_bright_2015}). The filter used for the 515 nm case was clearly less oxydized than others. 

Aluminum filters melt as soon as they are submitted to a few watts of average laser power. When HHG was driven by $3\omega_L$ or $4\omega_L$ beams, the average power was sufficiently low to preserve the filters. However for measurements using the fundamental or second harmonic, we need to remove a large part of the driving laser power before reaching the Al filters. For that purpose we used SiO$_2$ plates under 20$^\circ$ grazing incidence covered with a layer of Nb$_2$O$_5$.
These plates present a high transmission at the fundamental laser wavelength. The measured transmission for a set of two plates in s and p polarizaton are respectively up to 58 $\%$ and 99.3 $\%$ at 1030 nm and up  to 40$\%$ and 99.5 $\%$ at 515 nm.
We measured the VUV reflectivity of the plates by comparing the HHG spectrum reflected by two plates and sent directly to the spectrometer (Fig. \ref{fig5}). The XUV reflectivity for one plate lies in the 45$\%$ - 70$\%$ range in s polarization and  25 - 50$\%$ range in p polarization. 

To retrieve the XUV flux per harmonic, we used the total flux measured on the photodiode, the harmonic spectrum measured by the XUV spectrometer in the same conditions, and calibrated out the transmission of the Al filter and the reflectivity of the Nb$_2$O$_5$ plates when used. We also took into account the spatial clipping of the beams by the setup, to retrieve the flux produced by the source. We estimate the uncertainity in the flux measurement to $\pm 20\%$ because the spatial clipping and the typical fluctuation of the current of the photodiode. Note that we did not renormalize the signal to correct for partial reabsorption of the harmonics by the residual gas \cite{klas_table-top_2016}.

\section{HHG Optimization}

\begin{figure}[htbp]
\centering
\fbox{\includegraphics[width=\linewidth]{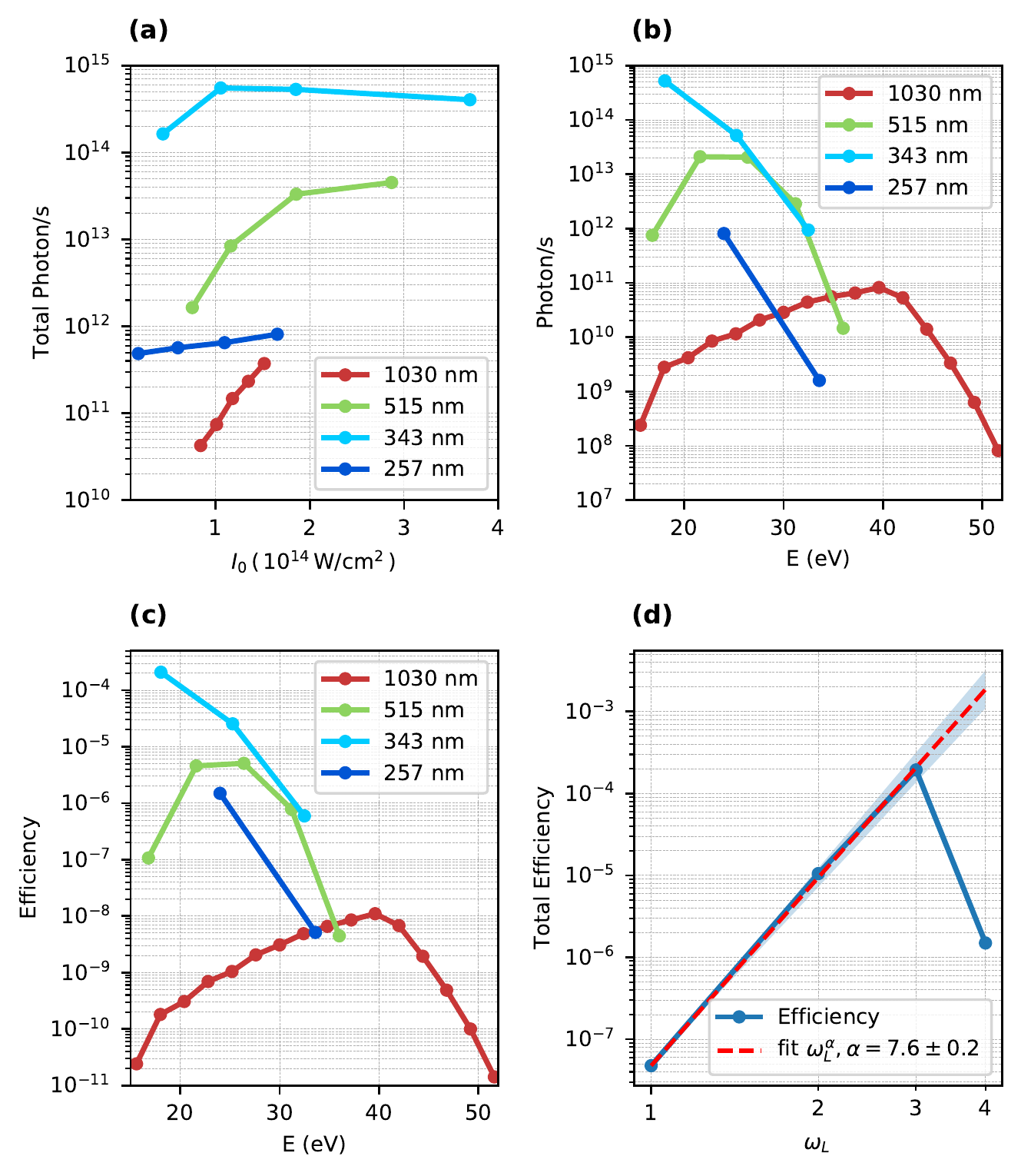}}
%\fbox{\includegraphics[width=\linewidth]{fig3}}
%\fbox{\input{fig3_pgf.pgf}}
\caption{(a) Intensity dependency of the high-harmonic flux in the optimized focusing conditions. The intensity was tuned by changing the laser energy. A tighter focusing would have lead to an higher intensity at the highest pulse energy but also to a lower flux.
(b) Maximal flux at each harmonic generated from 1,2,3 and 4$\omega_L$ beams. 
(c) Maximal efficiency of the HHG process for each harmonic at 1,2,3 and 4$\omega_L$.
(d) Total efficiency for the detected harmonics as a function of generating frequency, at fixed intensity around 2 $10^{14}$ W/cm$^2$.  The dashed line is a power fit of the efficiency.
}
\label{fig3}
\end{figure}

The optimal generating conditions are determined by three processes: single atom response (nonlinearly increasing with laser intensity), XUV reabsorption (increasing with gas density and medium length) and phase matching. 
 
Phase matching can be quantified by calculating the optimal medium density \cite{heyl_scale-invariant_2016,rothhardt_absorption-limited_2014}: 
\begin{equation}
d_{PM}= d_0 \frac{\lambda^2}{2 \pi^2\omega_0^2 \Delta \delta (1-\eta /\eta_{crit})}
\label{dPM}
\end{equation} 

with $d_0$ the density at one atmosphere equal to $2.5 \times 10^{19}$ at./cm$^3$, $\Delta \delta$ the difference of refractive index between the fundamental and the XUV, $\eta$ the ionization fraction and $\eta_{crit}$ the critical ionization. The perfect medium density is the one that is equal to $d_{PM}$. To perform HHG with low energy pulses, one needs to focus the beam in a tight focusing geometry with small focus waist $w_0$ to reach the intensity of I $\sim 10^{14}$ W/cm$^2$ level. The small $w_0$ leads to an high density jet by eq. \ref{dPM} and also a short length in order to avoid a too strong reabsorption. We thus designed and characterized a short and dense gas jet \cite{comby_absolute_2018} where we perform an absorption limited HHG. 

A crucial parameter to be optimized for the HHG is the intensity at the generation media, which plays a important role on the single-atom response and on the phase-matching through the ionization fraction. The atomic dipole response scales non-linearly with the laser intensity, with a typical effective nonlinearity of $q_{\mathrm{eff}}\sim 4$ in the plateau region. The harmonic flux is thus expected to scale as $\sim$ I$_L^{q_{\mathrm{eff}}}{\propto } \left( E_L / w_0^2\right)^{q_{\mathrm{eff}}}$ , where  $E_L$ is the laser pulse energy. It thus seems that the tighter we focus, the higher is the flux. However, this statement must be nuanced. If the intensity is too high, the atoms are fully ionized on the rising edge of the pulse and no more ground state population is available to produce HHG when the driving laser pulse intensity reaches its maximum. Moreover, a strong ionization could also lead to a very detrimental phase-matching due too high ionization fraction $\eta$ above $\eta_{crit}$. 

Our method of optimization of the high-harmonic flux consists of finding the right focus waist $w_0$ to have the optimal intensity that maximizes the XUV conversion efficiency. Experimentally, we proceeded as follow : we tuned the position of the telescope and the focal length of the lens to set $w_0$, then we increased step by step the pulse energy $E_L$ and measured the HHG flux at each step. If the maximum flux was not obtained at the maximum pulse energy, this means that we focused too tightly. We then decreased the $w_0$ until we reached the maximum flux at the maximum pulse energy. As the 3 and 4$\omega_L$ beam size changes with the power, we had to adapt our protocol. For these cases, we tuned the intensity by setting pinholes with different diameter in front of the beam at its maximum power. We measured the transmitted power for each pinhole diameter, and we estimate the size of the focal waist using basic fourier optics. Note that for every $\omega_L$, we also optimized all the other experimental parameters such as focus position, length and density of the Ar gas jet.

Generating high-harmonics from the fundamental 1030 nm laser beam with 300 $\mu$J, we found that the optimized focused conditions were obtained with a -100 mm/300 mm dielectric mirror telescope associated to a focal lens of 300 mm. This leads to a beam waist at focus $w_0 = 26\mu m$. With these conditions, the harmonic signal was found to increase with laser intensity $I_L$ (set by tuning the laser pulse energy), scaling as $I_L^4$ and reaching a maximum for  $I_L=1.5\times10^{14}$ W.cm$^{-2}$ (Fig. \ref{fig3}(a)). Further increasing the intensity required changing the focusing geometry and resulted in a lower signal at the maximum pulse energy. We measured the gas density profile of the jet and found a 4.5 $10^{18}$ at./cm$^3$ peak density and a length of 360 $\mu$m FWHM. 

For HHG from the 2-4$\omega_L$, the signals were optimal for lower gas densities, about $3.5 \times 10^{18}$ at./cm$^3$ with a medium length of 360 $\mu$m. At 2$\omega_L$, the best conditions were obtained with a -100 mm/300 mm dichroic mirror telescope and a f = 200 mm lens, leading to a beam waist at focus $w_0 =13 \mu m$. At low intensities, around $1.5\times 10^{14}$ W.cm$^{-2}$, the harmonic signal is almost two orders of magnitude higher than using the 1030 nm fundamental pulses (Fig. \ref{fig3}(a)).  The signal increases fast with laser intensity and saturates when $I_L$ reaches around 3$\times 10^{14}$ W.cm$^{-2}$. 
At 3$\omega_L$, we used a -100 mm/500 mm dichroic mirror telescope whose distance were tuned to get the proper magnification, followed by a f=300 mm lens, leading to $w_0 =8 \mu m$. The harmonic signal reaches the $10^{14}$ photons/s range and saturates when the intensity is between 1 and 3 $\times 10^{14}$ W.cm$^{-2}$ (Fig. \ref{fig3}(a)). Last, at 4$\omega_L$ we use also a -100 mm/500 mm telescop that leads to $\omega_1$  = 4 mm and we used focal lens of 200 mm. By assuming $DDL=1.4$ (same as 3$\omega_L$), we have $w_0 =5.5 \mu m$. The harmonic signal is more than two orders of magnitude lower than the one obtained at 3$\omega_L$.

We can be more quantitative about the phase-matching process by estimating the phase-matching density $d_{PM}$ and the absorption length. The absorption length at 27 eV photon energy in Ar is equal to 65 $\mu$m at $4.5 \times 10^{18}$ at./cm$^3$ (at $\omega_L)$, and is equal to 85 $\mu$m at $3.5 \times 10^{18}$ at./cm$^3$ (at 2,3 and 4$\omega_L)$ . With a jet length of 360 $\mu$m FWHM, we satisfy for each $\omega_L$  the first condition claiming that the medium should be at least three times higher than the absorption length \cite{constant_optimizing_1999}, a configuration where the harmonics are produced at the end of the jet. The phase-matching density $d_{PM}$ at $\omega_L$ is  $4 \times10^{18}$ at./cm$^3$ for a zero ionization yield ($\eta=0$), with $w_0 = 26\mu m$ and at 27 eV where $\Delta \delta \sim 7.1 \times 10^{-4}$  \cite{henke_x-ray_1993}. $d_{PM}$ is in the same range for 2, 3 and 4 $\omega_L$ and is respectively equal to  2.8, 3.3 and 4.0 $\times10^{18}$ at./cm$^3$ at zero ionization yield, which is close to the gas jet peak density. In order to estimate the exact $d_{PM}$, we calculated the ionization fraction by performing simulation using the Yudin-Ivanov model \cite{yudin_nonadiabatic_2001} and obtained $\eta\sim 2 \%$ at the middle of a pulse whose peak power is  $1 \times 10^{14}$ W.cm$^{-2}$, and $\eta\sim 20 \%$ for a $1.5 \times 10^{14}$ W.cm$^{-2}$ pulse. We thus obtain $d_{PM} \sim 8 \times10^{18}$ at./cm$^3$ for an intensity of $1 \times 10^{14}$ W.cm$^{-2}$. With a peak density of $4.5 \times 10^{18}$ at./cm$^3$ in the experiment, we are not far from the ultimate phase-matching pressure at this intensity. The peak intensity optimizing the HHG flux was however higher than this value ($1.5 \times 10^{14}$ W.cm$^{-2}$), see Fig. \ref{fig3}(a). This means that the phase-matching conditions were not perfectly fulfilled in the whole medium, and during the whole pulse duration : we have a transient phase-matching. The transient phase-matching is unfavorable to a good build-up of the XUV, but it is compensated by the increase of the single-atom response with increasing intensity. Moreover the efficient XUV build-up could also be done on a surface around the center of the pulse where the intensity is less high. Generally, one remarkable feature in the observations made in Fig. \ref{fig3}(a) is the rather high intensities at which the harmonic signals are optimized. Indeed we expect thanks to Yudin-Ivanov simulation a fully ionized medium at 2,3 and 4 $\omega_L$. A strong transient phase-matching is thus present at these laser frequencies. However, the transient phase-matching is less a problem at high $\omega_L$. Indeed, $\eta_{crit}$ increases with laser frequency, to $15,35$ and $45\%$ around 25 eV for 2,3 and 4$\omega_L$ respectively \cite{paul_phase-matching_2006}. We conclude that for each $\omega_L$ we maximise our flux under conditions where we have a strong transient phase matching that becomes less an issue at higher $\omega_L$. 

\section{HHG Flux and Efficiency}

To illustrate the versatility of this XUV source, we display the maximum flux of each harmonic in Fig. \ref{fig3} (b) from optimal HHG from 1-4$\omega_L$. At $\omega_L$, the XUV photon flux reaches almost the 10$^{11}$ photons/s around 35-40 eV. Switching to 2$\omega_L$ generating pulses increases the flux by two orders of magnitude, reaching $3.5 \times 10^{13}$ photons/s at 26.4 eV. The HHG process is even more efficient at 3$\omega_L$ where we produce $5.1 \times 10^{14}$ photons/s in a single harmonic. This brings the source to the mW level (1.5 mW average power at 166 kHz). This represents to the best of our knowledge the highest flux in the XUV range in a single pass system without post-compression stage \cite{klas_table-top_2016}. This flux is almost as high as the highest one reported in a cavity \cite{porat_phase-matched_2018}, but using a remarkably simple system. At 4$\omega_L$ the flux drops at $10^{11}$ photons/s but is still higher than the flux at $\omega_L$. One advantage of the 4$\omega_L$ is the large spacing between consecutive harmonics (9.6 eV), which makes their spectral selection for applications easier.

\begin{table}[t]
  \centering
\begin{tabular}{| c || c | c | c | c |}
\hline
 & Photon Energy (eV) & 25 & 39.6 & 44.4\\
\cline{2-5}
$\omega_L$ & Flux ($ \times 10^{10}$ Photon/s)  & 1.2$\pm$0.2 & 8.0$\pm$1.5 & 1.5$\pm$0.3 \\
\cline{2-5}
1030 nm & Power (nW) & 50$\pm$10 & 500$\pm$100 & 100$\pm$20 \\
\hline \hline

 & Photon Energy (eV) & 21.6 & 26.4 & 31.2 \\
\cline{2-5}
$2\omega_L$ & Flux ($ \times 10^{12}$ Photon/s)  & 21$\pm$4 & 20$\pm$4 & 3$\pm$0.5 \\
\cline{2-5}
515 nm & Power ($\mu$W) & 75$\pm$15 & 90$\pm$20 & 15$\pm$3 \\
\hline \hline

 & Photon Energy (eV) & 18 & 25.2 & - \\
\cline{2-5}
$3\omega_L$ & Flux ($ \times 10^{14}$ Photon/s)  & 5.1$\pm$1 & 0.5$\pm$0.1 & - \\
\cline{2-5}
343 nm & Power (mW) & 1.5$\pm$0.2 & 0.20$\pm$0.04 & - \\
\hline \hline

 & Photon Energy (eV) & 24 & - & - \\
\cline{2-5}
$4\omega_L$ & Flux ($ \times 10^{10}$ Photon/s)  & 80$\pm$20 & - & - \\
\cline{2-5}
257 nm & Power ($\mu$W) & 3$\pm$1 & - & - \\
\hline

\end{tabular}
\caption{Measured XUV photon flux at optimal conditions for the four driving wavelength. We present the most relevant photon energy.}
  \label{tab1}
\end{table}

An important characteristic of secondary XUV sources is their conversion efficiency, defined as the fraction of generating laser power converted in the XUV range. We plot in Fig. \ref{fig3} (d) the optimized efficiency for each high-order harmonic at each driving laser frequency. The efficiency for the $\omega_L$ is about $1 \times 10^{-8}$ in the 30-40 eV regime and gain orders of magnitude to $5 \times 10^{-6}$ at 2$\omega_L$ at 26.2 eV in argon. This value is lower compared to other sources, which could be due to residual reabsorption effects in our experiment \cite{harth_compact_2017}. It could have been also beneficial to use a shorter and denser gas jet to be more in agreement with the perfect density increasing at high ionization fraction. The best efficiency is obtained at 3$\omega_L$ (343 nm) and reaches about $2.0 \times 10^{-4}$ at 18 eV. This is above the value reported using 400 nm pulses ($5 \times 10^{-5}$) \cite{wang_bright_2015}, and slightly lower than the one measured from much shorter (25 fs) 270 nm pulses ($5 \times 10^{-4}$ at 22 eV) \cite{popmintchev_ultraviolet_2015}.

In order to evaluate the dependence of the harmonic yield on the generating laser frequency, we plot the efficiency of the HHG process, summed over all detected harmonics produced at an intensity around $1.5 - 2.0 \times 10^{14}$ W/cm$^2$, in Fig. \ref{fig3} (d). The evolution in log-log scale is linear when the generating frequency increases from $\omega_L$ to $3\omega_L$, and can be fitted as a $\omega_L^\alpha$ scaling with $\alpha = 7.6 \pm 0.2$ in a 2 $\sigma$ interval. This value seems a bit far from the expected $\lambda^{-5.5} \equiv \omega_L^{5.5}$ predicted by TDSE calculation of the atomic response \cite{colosimo_scaling_2008}. However we also have to take into account the propagation effect in a highly ionized medium where short wavelength are less perturbated than long wavelength. This increases the $\omega_L$ dependency, in agreement to the observations of Popmintchev et al. \cite{popmintchev_ultraviolet_2015} who expected a $\omega_L^{8}$ law in a highly ionized medium. This high dependency on $\omega_L$ underlines the benefit to generate VUV with shorter wavelength.

%In order to improve the efficiency we also perform experiment on Kr and Xe gas jet. But even by changing the focusing geometry to adapt to the lower ionization potential and lower over-barrier intensity, we do not succeed to reach better efficiency. We were close to the one in Ar, but our slightly lower efficiency may arise from the fact that we are at a lower working intensity so that we loose too much on the $I^{q_{\mathrm{eff}}}$ law of the dipole response. 

It is finally worthwhile to compare the efficiency from the raw power of the laser source. From our 50 W - 130 fs laser power we thus have an efficiency of $3.0\pm0.6 \times 10^{-5}$, which is higher than the $7 \times 10^{-6}$ from 120 W - 300 fs\cite{klas_table-top_2016} and $2.5 \times 10^{-5}$ from 80 W - 120 fs YDFA \cite{porat_phase-matched_2018} but with a much simpler experimental setup that could be handle easily.

\section*{Conclusion}

High order harmonics from harmonics provide an ideal source for applications in the 15-30 eV range. The simplicity of the setup and the very high conversion efficiencies open the route to a broad range of experiments in photoionization, photoemission, or transient absorption spectroscopy. 

This conclusion is however not valid at higher photon energy, since the cutoff from harmonics produced by $\omega_L$ is higher than the one produced from harmonics. The limiting factor to push the cutoff of the harmonic emission is here the laser pulse duration. In our conditions, calculations of the ionization fraction using the Yudin-Ivanov model \cite{yudin_nonadiabatic_2001} show that full ionization is reached before the maximum intensity of the laser pulse in the 1-2-3$\omega_L$ cases which means that the harmonics are only produced during a fraction of the pulse (the leading edge) and generating volume. Reducing the pulse duration from 130 to 15 fs \cite{lavenu_high-energy_2017} will allow to efficiently generate high order harmonic at the high effective laser intensity and reach even higher conversion efficiencies. Furthermore, these shorter pulse would yield higher cutoffs, typically in the 60-120 eV range, enabling important absorption edges to be reached.

\section*{Funding Information}
European Research Council (ERC 682978 - EXCITERS). French National Research Agency through (ANR-14-CE32-0014 MISFITS).

\section*{Acknowledgments}
The authors thank R. Bouillaud, L. Merzeau, N. Fedorov for technical support and  F. Catoire for fruitful discussions.

\bibliographystyle{acm}
\bibliography{w4wflux_OE_Arxiv.bbl}

\begin{thebibliography}{10}

\bibitem{balcou_optimizing_1992}
{\sc Balcou, P., Cornaggia, C., Gomes, A. S.~L., Lompre, L.~A., and
  L{\textbackslash}textquotesingleHuillier, A.}
\newblock Optimizing high-order harmonic generation in strong fields.
\newblock {\em Journal of Physics B: Atomic, Molecular and Optical Physics 25},
  21 (Nov. 1992), 4467--4485.

\bibitem{boullet_high-order_2009}
{\sc Boullet, J., Zaouter, Y., Limpert, J., Petit, S., Mairesse, Y., Fabre, B.,
  Higuet, J., Mével, E., Constant, E., and Cormier, E.}
\newblock High-order harmonic generation at a megahertz-level repetition rate
  directly driven by an ytterbium-doped-fiber chirped-pulse amplification
  system.
\newblock {\em Optics Letters 34}, 9 (May 2009), 1489--1491.

\bibitem{cabasse_optimization_2012}
{\sc Cabasse, A., Machinet, G., Dubrouil, A., Cormier, E., and Constant, E.}
\newblock Optimization and phase matching of fiber-laser-driven high-order
  harmonic generation at high repetition rate.
\newblock {\em Optics Letters 37}, 22 (2012), 4618--4620.

\bibitem{calegari_advances_2016}
{\sc Calegari, F., Sansone, G., Stagira, S., Vozzi, C., and Nisoli, M.}
\newblock Advances in attosecond science.
\newblock {\em Journal of Physics B: Atomic, Molecular and Optical Physics 49},
  6 (Mar. 2016), 062001.

\bibitem{colosimo_scaling_2008}
{\sc Colosimo, P., Doumy, G., Blaga, C.~I., Wheeler, J., Hauri, C., Catoire,
  F., Tate, J., Chirla, R., March, A.~M., Paulus, G.~G., Muller, H.~G.,
  Agostini, P., and DiMauro, L.~F.}
\newblock Scaling strong-field interactions towards the classical limit.
\newblock {\em Nature Physics 4}, 5 (May 2008), 386--389.

\bibitem{comby_absolute_2018}
{\sc Comby, A., Beaulieu, S., Constant, E., Descamps, D., Petit, S., and
  Mairesse, Y.}
\newblock Absolute gas density profiling in high-order harmonic generation.
\newblock {\em Optics Express 26}, 5 (Mar. 2018), 6001--6009.

\bibitem{constant_optimizing_1999}
{\sc Constant, E., Garzella, D., Breger, P., Mével, E., Dorrer, C., Le~Blanc,
  C., Salin, F., and Agostini, P.}
\newblock Optimizing {High} {Harmonic} {Generation} in {Absorbing} {Gases}:
  {Model} and {Experiment}.
\newblock {\em Physical Review Letters 82}, 8 (Feb. 1999), 1668--1671.

\bibitem{cooper_photoionization_1962}
{\sc Cooper, J.~W.}
\newblock Photoionization from {Outer} {Atomic} {Subshells}. {A} {Model}
  {Study}.
\newblock {\em Physical Review 128}, 2 (Oct. 1962), 681--693.

\bibitem{corkum_plasma_1993}
{\sc Corkum, P.~B.}
\newblock Plasma perspective on strong field multiphoton ionization.
\newblock {\em Physical Review Letters 71}, 13 (1993), 1994.

\bibitem{dubietis_two-photon_2000}
{\sc Dubietis, A., Tamošauskas, G., Varanavičius, A., and Valiulis, G.}
\newblock Two-photon absorbing properties of ultraviolet phase-matchable
  crystals at 264 and 211 nm.
\newblock {\em Applied Optics 39}, 15 (May 2000), 2437.

\bibitem{dorner_cold_2000}
{\sc Dörner, R., Mergel, V., Jagutzki, O., Spielberger, L., Ullrich, J.,
  Moshammer, R., and Schmidt-Böcking, H.}
\newblock Cold {Target} {Recoil} {Ion} {Momentum} {Spectroscopy}: a ‘momentum
  microscope’ to view atomic collision dynamics.
\newblock {\em Physics Reports 330}, 2 (June 2000), 95--192.

\bibitem{gagnon_time-resolved_2008}
{\sc Gagnon, E., Sandhu, A.~S., Paul, A., Hagen, K., Czasch, A., Jahnke, T.,
  Ranitovic, P., Lewis~Cocke, C., Walker, B., Murnane, M.~M., and Kapteyn,
  H.~C.}
\newblock Time-resolved momentum imaging system for molecular dynamics studies
  using a tabletop ultrafast extreme-ultraviolet light source.
\newblock {\em Review of Scientific Instruments 79}, 6 (June 2008), 063102.

\bibitem{harth_compact_2017}
{\sc Harth, A., Guo, C., Cheng, Y.-C., Losquin, A., Miranda, M., Mikaelsson,
  S., Heyl, C.~M., Prochnow, O., Ahrens, J., Morgner, U., L'Huillier, A., and
  Arnold, C.~L.}
\newblock Compact 200 {kHz} {HHG} source driven by a few-cycle {OPCPA}.
\newblock {\em Journal of Optics 20}, 1 (Dec. 2017), 014007.

\bibitem{henke_x-ray_1993}
{\sc Henke, B., Gullikson, E., and Davis, J.}
\newblock X-ray interactions: photoabsorption, scattering, transmission, and
  reflection at {E}=50-30000 {eV}, {Z}=1-92.
\newblock {\em Atomic Data and Nuclear Data Tables 54}, 2 (1993), 181--342.

\bibitem{heyl_introduction_2017}
{\sc Heyl, C.~M., Arnold, C.~L., Couairon, A., and L’Huillier, A.}
\newblock Introduction to macroscopic power scaling principles for high-order
  harmonic generation.
\newblock {\em Journal of Physics B: Atomic, Molecular and Optical Physics 50},
  1 (Jan. 2017), 013001.

\bibitem{heyl_scale-invariant_2016}
{\sc Heyl, C.~M., Coudert-Alteirac, H., Miranda, M., Louisy, M., Kovacs, K.,
  Tosa, V., Balogh, E., Varjú, K., L’Huillier, A., Couairon, A., and Arnold,
  C.~L.}
\newblock Scale-invariant nonlinear optics in gases.
\newblock {\em Optica 3}, 1 (Jan. 2016), 75.

\bibitem{heyl_high-order_2012}
{\sc Heyl, C.~M., Güdde, J., L’Huillier, A., and Höfer, U.}
\newblock High-order harmonic generation with uj laser pulses at high
  repetition rates.
\newblock {\em Journal of Physics B: Atomic, Molecular and Optical Physics 45},
  7 (Apr. 2012), 074020.

\bibitem{higuet_high-order_2011}
{\sc Higuet, J., Ruf, H., Thiré, N., Cireasa, R., Constant, E., Cormier, E.,
  Descamps, D., Mével, E., Petit, S., Pons, B., Mairesse, Y., and Fabre, B.}
\newblock High-order harmonic spectroscopy of the {Cooper} minimum in argon:
  {Experimental} and theoretical study.
\newblock {\em Physical Review A 83}, 5 (May 2011).

\bibitem{hadrich_single-pass_2016}
{\sc Hädrich, S., {Jan Rothhardt}, Krebs, M., Demmler, S., Klenke, A.,
  Tünnermann, A., and Limpert, J.}
\newblock Single-pass high harmonic generation at high repetition rate and
  photon flux.
\newblock {\em Journal of Physics B: Atomic, Molecular and Optical Physics 49},
  17 (Sept. 2016), 172002.

\bibitem{hadrich_energetic_2016}
{\sc Hädrich, S., Kienel, M., Müller, M., Klenke, A., Rothhardt, J., Klas,
  R., Gottschall, T., Eidam, T., Drozdy, A., Jójárt, P., Várallyay, Z.,
  Cormier, E., Osvay, K., Tünnermann, A., and Limpert, J.}
\newblock Energetic sub-2-cycle laser with 216 {W} average power.
\newblock {\em Optics Letters 41}, 18 (Sept. 2016), 4332--4335.

\bibitem{kaumanns_multipass_2018}
{\sc Kaumanns, M., Pervak, V., Kormin, D., Leshchenko, V., Kessel, A., Ueffing,
  M., Chen, Y., and Nubbemeyer, T.}
\newblock Multipass spectral broadening of 18 {mJ} pulses compressible from 13
  ps to 41 fs.
\newblock {\em Optics Letters 43}, 23 (Dec. 2018), 5877.

\bibitem{kazamias_pressure-induced_2011}
{\sc Kazamias, S., Daboussi, S., Guilbaud, O., Cassou, K., Ros, D., Cros, B.,
  and Maynard, G.}
\newblock Pressure-induced phase matching in high-order harmonic generation.
\newblock {\em Physical Review A 83}, 6 (June 2011).

\bibitem{klas_table-top_2016}
{\sc Klas, R., Demmler, S., Tschernajew, M., Hädrich, S., Shamir, Y.,
  Tünnermann, A., Rothhardt, J., and Limpert, J.}
\newblock Table-top milliwatt-class extreme ultraviolet high harmonic light
  source.
\newblock {\em Optica 3}, 11 (Nov. 2016), 1167.

\bibitem{lai_wavelength_2013}
{\sc Lai, C.-J., Cirmi, G., Hong, K.-H., Moses, J., Huang, S.-W., Granados, E.,
  Keathley, P., Bhardwaj, S., and Kärtner, F.~X.}
\newblock Wavelength {Scaling} of {High} {Harmonic} {Generation} {Close} to the
  {Multiphoton} {Ionization} {Regime}.
\newblock {\em Physical Review Letters 111}, 7 (Aug. 2013), 073901.

\bibitem{lavenu_nonlinear_2018}
{\sc Lavenu, L., Natile, M., Guichard, F., Zaouter, Y., Delen, X., Hanna, M.,
  Mottay, E., and Georges, P.}
\newblock Nonlinear pulse compression based on a gas-filled multipass cell.
\newblock {\em Optics Letters 43}, 10 (May 2018), 2252.

\bibitem{lavenu_high-energy_2017}
{\sc Lavenu, L., Natile, M., Guichard, F., Zaouter, Y., Hanna, M., Mottay, E.,
  and Georges, P.}
\newblock High-energy few-cycle {Yb}-doped fiber amplifier source based on a
  single nonlinear compression stage.
\newblock {\em Optics Express 25}, 7 (Apr. 2017), 7530.

\bibitem{lewenstein_theory_1994}
{\sc Lewenstein, M., Balcou, P., Ivanov, M.~Y., L’huillier, A., and Corkum,
  P.~B.}
\newblock Theory of high-harmonic generation by low-frequency laser fields.
\newblock {\em Physical Review A 49}, 3 (1994), 2117.

\bibitem{marceau_wavelength_2017}
{\sc Marceau, C., Hammond, T.~J., Naumov, A.~Y., Corkum, P.~B., and Villeneuve,
  D.~M.}
\newblock Wavelength scaling of high harmonic generation for 267 nm, 400 nm and
  800 nm driving laser pulses.
\newblock {\em Journal of Physics Communications 1}, 1 (Sept. 2017), 015009.

\bibitem{paul_phase-matching_2006}
{\sc Paul, A., Gibson, E., Zhang, X., Lytle, A., Popmintchev, T., Zhou, X.,
  Murnane, M., Christov, I., and Kapteyn, H.}
\newblock Phase-{Matching} {Techniques} for {Coherent} {Soft} {X}-{Ray}
  {Generation}.
\newblock {\em IEEE Journal of Quantum Electronics 42}, 1 (Jan. 2006), 14--26.

\bibitem{paul_observation_2001}
{\sc Paul, P.~M., Toma, E.~S., Breger, P., Mullot, G., Augé, F., Balcou, P.,
  Muller, H.~G., and Agostini, P.}
\newblock Observation of a {Train} of {Attosecond} {Pulses} from {High}
  {Harmonic} {Generation}.
\newblock {\em Science 292}, 5522 (June 2001), 1689--1692.

\bibitem{popmintchev_ultraviolet_2015}
{\sc Popmintchev, D., Hernández-García, C., Dollar, F., Mancuso, C.,
  Pérez-Hernández, J.~A., Chen, M.-C., Hankla, A., Gao, X., Shim, B., Gaeta,
  A.~L., Tarazkar, M., Romanov, D.~A., Levis, R.~J., Gaffney, J.~A., Foord, M.,
  Libby, S.~B., Jaron-Becker, A., Becker, A., Plaja, L., Murnane, M.~M.,
  Kapteyn, H.~C., and Popmintchev, T.}
\newblock Ultraviolet surprise: {Efficient} soft x-ray high-harmonic generation
  in multiply ionized plasmas.
\newblock {\em Science 350}, 6265 (Dec. 2015), 1225--1231.

\bibitem{porat_phase-matched_2018}
{\sc Porat, G., Heyl, C.~M., Schoun, S.~B., Benko, C., Dörre, N., Corwin,
  K.~L., and Ye, J.}
\newblock Phase-matched extreme-ultraviolet frequency-comb generation.
\newblock {\em Nature Photonics 12}, 7 (July 2018), 387--391.

\bibitem{puppin_time-_2019}
{\sc Puppin, M., Deng, Y., Nicholson, C.~W., Feldl, J., Schröter, N. B.~M.,
  Vita, H., Kirchmann, P.~S., Monney, C., Rettig, L., Wolf, M., and Ernstorfer,
  R.}
\newblock Time- and angle-resolved photoemission spectroscopy of solids in the
  extreme ultraviolet at 500 {kHz} repetition rate.
\newblock {\em Review of Scientific Instruments 90}, 2 (Feb. 2019), 023104.

\bibitem{rothhardt_53w_2014}
{\sc Rothhardt, J., Hädrich, S., Klenke, A., Demmler, S., Hoffmann, A.,
  Gotschall, T., Eidam, T., Krebs, M., Limpert, J., and Tünnermann, A.}
\newblock 53w average power few-cycle fiber laser system generating soft x rays
  up to the water window.
\newblock {\em Optics Letters 39}, 17 (Sept. 2014), 5224--5227.

\bibitem{rothhardt_high-repetition-rate_2016}
{\sc Rothhardt, J., Hädrich, S., Shamir, Y., Tschnernajew, M., Klas, R.,
  Hoffmann, A., Tadesse, G.~K., Klenke, A., Gottschall, T., Eidam, T., Limpert,
  J., Tünnermann, A., Boll, R., Bomme, C., Dachraoui, H., Erk, B., Di~Fraia,
  M., Horke, D.~A., Kierspel, T., Mullins, T., Przystawik, A., Savelyev, E.,
  Wiese, J., Laarmann, T., Küpper, J., and Rolles, D.}
\newblock High-repetition-rate and high-photon-flux 70 {eV} high-harmonic
  source for coincidence ion imaging of gas-phase molecules.
\newblock {\em Optics Express 24}, 16 (Aug. 2016), 18133.

\bibitem{rothhardt_absorption-limited_2014}
{\sc Rothhardt, J., Krebs, M., Hädrich, S., Demmler, S., Limpert, J., and
  Tünnermann, A.}
\newblock Absorption-limited and phase-matched high harmonic generation in the
  tight focusing regime.
\newblock {\em New Journal of Physics 16}, 3 (2014), 033022.

\bibitem{shiner_wavelength_2009}
{\sc Shiner, A.~D., Trallero-Herrero, C., Kajumba, N., Bandulet, H.-C.,
  Comtois, D., Légaré, F., Giguère, M., Kieffer, J.-C., Corkum, P.~B., and
  Villeneuve, D.~M.}
\newblock Wavelength {Scaling} of {High} {Harmonic} {Generation} {Efficiency}.
\newblock {\em Physical Review Letters 103}, 7 (Aug. 2009), 073902.

\bibitem{ueffing_nonlinear_2018}
{\sc Ueffing, M., Reiger, S., Kaumanns, M., Pervak, V., Trubetskov, M.,
  Nubbemeyer, T., and Krausz, F.}
\newblock Nonlinear pulse compression in a gas-filled multipass cell.
\newblock {\em Optics Letters 43}, 9 (May 2018), 2070.

\bibitem{wang_bright_2015}
{\sc Wang, H., Xu, Y., Ulonska, S., Robinson, J.~S., Ranitovic, P., and Kaindl,
  R.~A.}
\newblock Bright high-repetition-rate source of narrowband extreme-ultraviolet
  harmonics beyond 22 {eV}.
\newblock {\em Nature Communications 6\/} (June 2015), 7459.

\bibitem{yudin_nonadiabatic_2001}
{\sc Yudin, G.~L., and Ivanov, M.~Y.}
\newblock Nonadiabatic tunnel ionization: {Looking} inside a laser cycle.
\newblock {\em Physical Review A 64}, 1 (June 2001), 013409.

\end{thebibliography}
%\bibliography{bibzotero_3}
\end{document}